\documentclass[onecolumn,showpacs,nobibnotes,nofootinbib]{revtex4}
\usepackage{amsmath,amssymb,bm}
\usepackage{epsfig}
\usepackage{graphicx}
\usepackage{slashed}

\def\b{{\boldsymbol b}}

\def\x{{\boldsymbol x}}
\def\y{{\boldsymbol y}}

\def\0{{\boldsymbol 0}}

\def\xB{x_{\rm{Bj}}}
\def\Yg{Y_{\rm{gap}}}
 \def\simge{\mathrel{%
    \rlap{\raise 0.511ex \hbox{$>$}}{\lower 0.511ex \hbox{$\sim$}}}}
\def\simle{\mathrel{
   \rlap{\raise 0.511ex \hbox{$<$}}{\lower 0.511ex \hbox{$\sim$}}}}

 \long\def\comment#1{ }
\newcommand{\beqa}{\begin{eqnarray}}
\newcommand{\eeqa}{\end{eqnarray}}

    \newcommand{\rmd}{{\rm d}}   %ELS%
  \newcommand{\dif}{{\rm d}}
  \newcommand{\dY}{\dif Y}

  \newcommand{\del}{\partial}

  \newcommand{\mcal}{\mathcal}

  \newcommand{\rmL}{{\rm L}}

  \newcommand{\rmR}{{\rm R}}

  \newcommand{\nn}{\nonumber\\}

  \newcommand{\be}{\begin{equation}}
  \newcommand{\ee}{\end{equation}}

  \newcommand{\beq}{\begin{eqnarray}}
  \newcommand{\eeq}{\end{eqnarray}}
  \newcommand{\avg}[1]{\left\langle #1 \right\rangle}

\begin{document}
\begin{flushright}
CERN-PH-TH/2012-245
\par\end{flushright}

\title{\bf Running coupling and pomeron loop effects on inclusive and diffractive DIS cross sections}

\author{M.~B.~Gay~Ducati${}^{(1,2)}$,
E. G. de Oliveira${}^{3}$, J. T. de Santana Amaral${}^4$}
\affiliation{${}^1$Instituto de F\'{\i}sica, Universidade Federal
do Rio Grande do Sul, Caixa Postal 15051, 91501-970 - Porto Alegre, RS, Brazil\\
${}^2$ CERN, PH-TH, CH-1211, Geneva 23, Switzerland\\
${}^3$Instituto de F\'{\i}sica, Universidade de S\~{a}o Paulo, C.P. 66318,
05315-970 S\~{a}o Paulo, SP, Brazil\\
${}^4$Instituto de F\'{\i}sica e Matem\'atica,  Universidade Federal de Pelotas, Caixa Postal 354, CEP 96010-900, Pelotas, RS, Brazil
}

\begin{abstract}
Within the framework of a (1+1)--dimensional model which mimics high
energy QCD, we study the behavior of the cross sections for
inclusive and diffractive deep inelastic $\gamma^*h$ scattering cross sections.
We analyze the cases of both fixed and running coupling
within the mean field approximation, in which the evolution of the
scattering amplitude is described by the Balitsky-Kovchegov equation, and also
through the pomeron loop equations, which include in the evolution the
gluon number fluctuations. In the diffractive case, similarly to 
the inclusive one, the suppression of the diffusive scaling, as a consequence of the inclusion of the running of the coupling, is observed.

\end{abstract}


\maketitle

\section{Introduction}\label{sec:intro}

It is well known that the high energy regime of the 
Quantum Chromodynamics (QCD) is described by non--linear evolution
equations \cite{GLR,MQ85,AM90,AGL-1,AGL-2,AGL-3,JKLW,JKLW1,JKLW2,JKLW3,
CGC,CGC1,CGC2,SAT,AB01,Balitsky,Balitsky2,B1,W}. At the level of scattering
amplitudes, and in the framework of the dipole picture \cite{AM94,AMPatel,AM95}, the most general ones are the so called
\textit{pomeron loop equations} \cite{IT04,MSW05,IT05,LL05}, which
correspond to a generalization of the Balitsky-JIMWLK hierarchy
\cite{Balitsky,Balitsky2,B1,JKLW,JKLW1,JKLW2,JKLW3,CGC,CGC1,CGC2,SAT,W},
by including the \textit{gluon number fluctuations}. If one performs a
mean field approximation, this infinite set of equations reduces to a single
closed equation for the scattering amplitude of one dipole with a hadronic target, the Balitsky-Kovchegov (BK) equation \cite{Balitsky,K99a,K99b}, the simplest of the non--linear equations
for the scattering amplitudes in QCD at high energy. This equation admits
\cite{MP03,MP03-1,MP03-2} travelling wave solutions, which have become a
natural explanation for the \textit{geometric scaling}--- first observed in
the HERA data for electron--proton deep inelastic scattering
\cite{geometric,MS06}--- and, being a mean field version of the complete
hierarchy, neglects the effects of the fluctuations. At least in the fixed coupling case, from the correspondence between
high energy QCD and reaction diffusion processes, one of the consequences
of the gluon number fluctuations in the evolution of the dipole scattering amplitudes is, at very high energies, the replacement of the \textit{geometric
scaling} \cite{geometric,MS06}, by the \textit{diffusive scaling} \cite{HIMST06}.

Fluctuation effects have not been observed in the experimental
data yet. Besides, the only few phenomenological studies have been inconclusive with respect to their presence in the current experiments
\cite{Kozlov:2007wm,agbs-fluct,F11,GA2012}. Their physical consequences in the high--energy evolution in QCD for the phenomenology were first analyzed in Ref.\
\cite{HIMST06}, where their effects in the behaviour of inclusive and diffractive cross sections for deep inelastic lepton--hadron scattering (DIS) were studied. They found, for example, that, within the high energy regime, all the amplitudes or cross sections show diffusive scaling, that is, they depend upon the photon virtuality $Q^2$ and the total rapidity $Y$ through the
variable $\ln(Q^2/\avg{Q_s^2})$, where $\avg{Q_s^2}$ is the (average) hadron saturation momentum.

Our current knowledge on the consequences of the fluctuations comes
only from the correspondence between high energy QCD and statistical
physics; because of the complexity of the pomeron loop equations,
the properties of the solutions are known only after some
approximations, in asymptotic regimes and at fixed coupling \cite{IT04}.
On the other hand, in the last few years one observed an important
progress in the inclusion of next-to-leading order (NLO) effects
in the non-linear mean field BK equation. In particular, one can
cite the explicit calculation of the running coupling effects
 \cite{kovwei1,kovwei2,javier_kov,balnlo,balnlo2,kovwei3,Kovchegov:2011aa} and
its successful use in the description of HERA and RHIC data
\cite{Albacete:2009fh,BGSA09,Albacete:2010sy}.
Unfortunately, because of the complexity of the pomeron loop equations,
the inclusion of such NLO effects in these equations turn to be a very
hard task. The difficulty of dealing with these equations, even in the
fixed coupling case, inspired other ways of investigation of high energy evolution in QCD, in particular through particle models with a smaller number
of dimensions \cite{Stasto05,KL5,SX05,KozLev06,BIT06,onedim,Dumitru:2007ew,Munier:2008cg}.
Among them, the (1+1)-dimensional model presented in
Ref.\cite{onedim} has shown to mimic fixed impact parameter high energy QCD
with fixed coupling constant. Its generalization to the
case with the running coupling was done in Ref. \cite{Dumitru:2007ew}. In
such version, the model could provide, for the first time, the study
of both running coupling and fluctuations effects, taken into account
simultaneously, in the high energy evolution of scattering amplitudes.
The main conclusion presented by the authors was the strong suppression of
the pomeron
loop (fluctuation) effects due to the running of the coupling, up to rapidity
$Y\simeq 200$, that is, well beyond the energies of interest for
the phenomenology in QCD. The dynamics is similar to the respective
prediction of the mean-field approximation with running coupling,
the property of (approximate) geometric scaling being preserved
for the average scattering amplitude. This result is in sharp
contrast with the fixed coupling results, which show the emergence
of the diffusive scaling.

In this paper we present an investigation of the effects
of both pomeron loops and running coupling, taken into account
simultaneously, on the cross sections for 
inclusive  $\gamma^*h$ and, for the first time, on diffractive deep inelastic scattering (DIS),
within the framework of the toy model \cite{Dumitru:2007ew}. In Sections \ref{sec:inclusive} 
and \ref{sec:diffractive} we present some important aspects of lepton-hadron DIS, specifically an overview of kinematics and the description of the dipole picture of the inclusive and
diffractive $\gamma^*h$ scattering. Section \ref{sec:model} is devoted to an overview of the
one-dimensional model in the running coupling case. In particular,
we present the resulting evolution equations for the scattering
amplitudes and the main features of their evolution. In Section 
\ref{sec:results} we present our results, with the study of the behaviour of the cross sections in both fixed and running coupling cases, and the conclusions are presented in Section \ref{sec:conclusions}.

\section{Inclusive virtual photon-hadron DIS}\label{sec:inclusive}

This process is described by the reaction $l(k)+h(P)\rightarrow l(k^\prime)+X(P_X)$,
where $l$ refers to the lepton (with momentum $k$ in the initial state and
$k^ \prime$ in the final one), $h$ to the incoming hadron (with momentum $P$)
and $X$ is the generic hadronic final state (with momentum $P_X$). Processes
described by the reaction above are called $inclusive$, because only the lepton
is measured in the final state. In the specific case where the lepton is an electron, its interaction with the hadron is mediated by a virtual photon with virtuality $Q^2=-q^ 2=(k-k^\prime)^ 2$. If one looks at the $\gamma^*h\rightarrow X$, in inclusive DIS all what is known from the final hadronic state $X$ is that it has an invariant mass squared $W^2=(P+q)^2$, which is the center-of-mass energy of the $\gamma^*h$ system. Another important definition is that of the Bjorken variable, or Bjorken-$x$, given by $\xB\equiv Q^2/(Q^2+W^2)$; from it, one sees that, for fixed values of $Q^2$, when one increases the energy $W^2$, $\xB$ decreases and the high energy limit corresponds to the small-$\xB$ limit. The total \textit{rapidity} of the process is defined as $Y\equiv \ln(1/\xB)$.

At small-$\xB$, the $\gamma^*h$ process can be described in a convenient
frame, the so-called \textit{dipole frame}, in which the hadron carries
most of the total energy, but the virtual photon has enough energy to
split into a quark-antiquark ($q\bar q$) pair, or a \textit{dipole}.
This dipole, then, interacts with the hadron. The dissociation of the
virtual photon into the color dipole takes place
long before the scattering, and the dipole evolves through soft gluon
radiation until it meets the hadron (at the time of scattering) and scatters
off the color fields therein. Exactly as it was done in \cite{HIMST06}, the
present analysis will be restricted to the leading logarithm approximation,
in which the evolution consists of the emission of soft gluons, carrying a small fraction $\xB\ll 1$ of the longitudinal momentum
of their parent parton. In the limit $N_c\rightarrow\infty$, a gluon can
be effectively replaced by a pointlike quark--antiquark pair in a color
octet state, and a soft gluon emission from a color dipole can be described
as the splitting of the original dipole into two new dipoles with a common
leg. In this picture, the original $q\bar q$ pair produced by the dissociation
of the virtual photon evolves through successive dipole splittings and becomes
an \textit{onium}---i.e., a collection of dipoles---at the time of scattering. This is the Mueller's dipole picture \cite{AM94,AMPatel,AM95}.

Using the formalism developed in \cite{HIMST06}, one finds that the differential cross-section for \textit{onium-hadron} scattering at fixed impact
parameter is given by
 \be\label{sigmab}
 \frac{\rmd\sigma_{\rm tot}}
 {\rmd^2 {b}}\,(\bm{r}, \bm{b}, Y)\,=\,2\,\textrm{Re}\,{\cal A}(\x,\y; Y),
 \ee
where ${\cal A}$ is the amplitude for the elastic scattering, $\bm{b}=(\x+\y)/2$
and $\bm{r}$ are the impact parameter and the transverse size of the original
dipole and $\x$ and $\y$ its transverse coordinates.

In such high energy approximation, the DIS cross section for
the inclusive \textit{virtual photon--hadron} ($\gamma^* h$) scattering can
be expressed as
 \be\label{eq:sigmatot}
 \frac{\rmd\sigma^{\gamma}_{\rm tot}}
 {\rmd^2 {b}}\,(Y,Q^2)
 =\int_0^1 \rmd v\int \rmd^2 {\bm r}\,\sum_{\alpha=L,T}
 \vert \psi^{\gamma}_{\alpha}(r, v; Q^2)\vert^2 \,
 2\,\textrm{Re}\,{\cal A}(\x,\y; Y), \ee
where $\vert \psi^{\gamma}_{T/L}\vert^2$ are the probability
densities for the $q \bar q$ dissociation of a virtual photon with transversal
($T$) or longitudinal ($L$) polarization, obtained from perturbative QED
\cite{AM90,NZ91}, given by
\begin{eqnarray}\label{eq:wavef}
\vert \Psi_{T}(r,v;Q^2) \vert^2
  & =& \frac{2N_c\alpha_{em}}{4\pi^2}\sum_{q}e^2_q
       \left\{\left[v^2+(1-v)^2\right]\bar Q_q^2K^2_1(\bar{Q}_q r)+m^2_qK^2_0(\bar{Q}_q r)\right\}\\
\vert \Psi_{L}(r,v;Q^2) \vert^2
  & =& \frac{2N_c\alpha_{em}}{4\pi^2}\sum_qe^2_q
       \left\{4Q^2v^2(1-v)^2K^2_0(\bar{Q}_q r)\right\},
\end{eqnarray}
where $\bar{Q}_q=v(1-v)Q^2+m^2_q$, $m_q$ is the mass of the quark with flavour
$q$, $K_{0,1}$ are the Mc Donald functions of rank zero and one, respectively,
and $v$ is the fraction of the photon longitudinal momentum carried
by the quark.

Expression (\ref{eq:sigmatot}) is a priori frame-independent,
but, the inclusive cross section is most simply evaluated in the frame
where almost all the total rapidity $Y$ is carried by the hadron (the target)
and the projectile is an elementary dipole. In this case,
${\cal A}(\x,\y; Y)=\avg{T(\x,\y)}_Y$ \cite{HIMST06} and
	 \be\label{eq:sigmatot_1}
		 \frac{\rmd\sigma^{\gamma}_{\rm tot}}
 		{\rmd^2 {b}}\,(Y,Q^2)
		 =\int_0^1 \rmd v\int \rmd^2 {\bm r}\,\sum_{\alpha=L,T}
		 \vert \psi^{\gamma}_{\alpha}(v, r; Q)\vert^2 \,
		2{\rm Re}\avg{T(\x,\y)}_Y,
\ee
where $\avg{T(\x,\y)}_Y$ is the (average) one dipole-hadron scattering amplitude, the brackets meaning the average over the target configurations.
Here, we are interested in the high-energy limit of the DIS cross sections at
fixed impact parameter. We assume that the dependence on $\b$ can be factorized into a profile function $S(\b)$, according to $\avg{T(\x,\y)}_Y=S(\b)\avg{T(r)}_Y$ ($r=|\x-\y|$ is the dipole size), where the integral $\sigma_0\!\equiv\!\int d\b\, S(\b)$ would provide an overall normalization factor of order of the transverse area of proton. Since the dependence on $\b$ in such an approximation results completely decoupled, in the following we simply set $S(\b)=1$, assuming the integration over $\b$ extended up to $b_{\rm max}$ providing the correct normalization of the cross section. The differential inclusive cross section reads
	 \be\label{eq:sigmatot_2}
		 \frac{\rmd\sigma^{\gamma}_{\rm tot}}
 		{\rmd^2 {b}}\,(Y,Q^2)
		 =4\pi\int_0^1 \rmd v\int_0^\infty \rmd r\,\sum_{\alpha=L,T}
		 \vert \psi^{\gamma}_{\alpha}(v, r; Q)\vert^2 \,
		r\,{\rm Re}\avg{T(r)}_Y.
	\ee
As it will be convenient for our purposes, we can write $\avg{T(r)}_Y\equiv
\avg{T(x)}_Y\equiv\avg{T_x}_Y$, where $x\equiv \ln(1/r^2Q_0^2)$ represents $r$ in logarithmic units\footnote{The
variable $x$ should not be confused with the bold--faced $\x$, which represents a vector in the transverse plane in
the picture of DIS.} ($Q_0$ is a scale of reference introduced by the initial conditions at
low energy). The total cross section, then, takes the form
	 \be\label{eq:sigmatot_3}
		 \frac{\rmd\sigma^{\gamma}_{\rm tot}}
 		{\rmd^2 {b}}\,(Y,Q^2)
		 =\frac{2\pi}{Q_0^2}\int_0^1 \rmd v\int_{-\infty}^{+\infty} \rmd x\,e^{-x}\sum_{\alpha=L,T}
		 \vert \psi^{\gamma}_{\alpha}(v, x; Q)\vert^2 \,
		{\rm Re}\avg{T_x}_Y.
	\ee

\section{Diffractive DIS}\label{sec:diffractive}

Part of the DIS events are diffractive. In such events, described by the reaction
$\gamma^*h\rightarrow Xh$, the final states contain an intact scattered hadron
$h$ and a diffractive hadronic state $X$ separated by a \textit{rapidity gap}
$\Yg\equiv \ln(1/x_\mathbb P)$, where $x_\mathbb P = \xB /\beta$ and
$\beta$ is related to the diffractive invariant mass $M_X$ by
$\beta\equiv Q^2/(Q^2+M_X^2)$. It is straightforward to see that the difference
between the total rapidity $Y$ and the rapidity gap $\Yg$ is $Y-\Yg=\ln(1/\beta)$.

The cross section for the diffractive process reads ($Y_{\rm gap}$ denotes the
minimal rapidity gap)
 \be\label{eq:sigmadiff}
 \frac{\rmd\sigma^{\gamma}_{\rm diff}}
 {\rmd^2 {b}}\,(Y,Y_{\rm gap},Q^2)
 =\int_0^1 \rmd v \int \rmd^2 {\bm r}\,\sum_{\alpha=L,T}
 \vert \psi^{\gamma}_{\alpha}(v, r; Q)\vert^2 \,
 P_{\rm diff}({\bm b}, {\bm r}; Y, Y_{\rm gap}).
 \ee
Since the whole process can be factorized, for our purposes it will be enough to
ignore the electomagnetic process (the splitting the virtual photon into
the $q\bar{q}$ dipole) and focus only on the onium--hadron (${\cal O}h$)
scattering. More specifically, we will be interested in the quantity $P_{\rm diff}$, which is the probability for diffractive onium-hadron scattering (${\cal O}h\to Xh$), and corresponds to the differential cross-section for onium-hadron scattering at fixed impact parameter:
 \be\label{sigma-oh}
 \frac{\rmd\sigma_{\rm diff}}
 {\rmd^2 {b}}\,(\bm{r}, \bm{b}, Y, \Yg)\,=\,
 P_{\rm diff}(\x,\y; Y, \Yg).
 \ee

An explicit formula for this probability has been obtained within the lightcone
wavefunction formalism in \cite{HIMST06}, in a special frame, in which $\Yg$ coincides with the rapidity $Y_0$ of the target hadron. This choice of the frame is important because it avoids one to deal explicitly with final state interactions. The resulting formula is given by
 \begin{eqnarray}\label{eq:Pdiff}
	 P_{\rm diff}(\x,\y; Y,\Yg)\rightarrow P_{\rm diff}(\x,\y; Y,Y_0)=
	\sum_{\{N\}}
 	P(\{N\}; Y-Y_0)
	\left|\avg{1-\prod_{i=1}^{N} S_{x_i}}_{Y_0}\right|^2 \, ,
\end{eqnarray}
which has the following meaning: starting from an original dipole $(\x,\y)$,
after an evolution $Y-Y_0$ there is a probability density 
$P(\{N\}; Y-Y_0)$ for a given configuration
of $N$ dipoles to be produced. $S_{x_i}= 1-T_{x_i}$ is the $S$--matrix for the scattering between the $i$th dipole (with logarithmic size $x_i$)
in the projectile and a given configuration of the target, $T_{x_i}$ being the corresponding $T$--matrix. The symbol $\sum_{\{N\}}$ represents the sum
over all the configurations of the projectile with $N$ dipoles. Again, the notation $\avg{\cdot}_{Y_0}$ denotes the average over the ensemble of color fields in the target.

Our main aim is to investigate, for the first time, the behavior of the diffractive probability (\ref{eq:Pdiff}) with increasing rapidity $Y$ in the presence of fluctuations and running coupling effects. This requires the
description of the rapidity evolution of the dipole--hadron
scattering amplitude $T_x$, as well as the probability density
$P(\{N\}; Y-Y_0)$ by taking into account both effects simultaneously, which
is still a prohibitive task in full QCD. However, a convenient way of doing this is through the model presented in \cite{Dumitru:2007ew}, whose main features we will briefly describe below.

\section{(1+1)--dimensional model for high energy QCD}\label{sec:model}

The toy model \cite{Dumitru:2007ew} is a (1+1)-dimensional stochastic particle model, where one of the dimensions refers to the total rapidity separation $Y$ between
two hadronic systems which undergo evolution and scattering (and plays
the role of time in the evolution), while the other one (the spatial dimension)
is the position of the particle along an infinite one-dimensional axis, the $x$-axis, which, in analogy with the dipole picture of QCD \cite{AM94,AM95}, corresponds to the logarithm of the inverse size of a dipole, as defined in
Section II.

\subsection{The structure of model}

In this toy model, a system of particles (which corresponds to a given hadronic
system) is specified by their distributions along the one-dimensional $x$-axis. In order to describe a scattering
problem, one considers two such systems (projectile and target)
which scatter off each other along
a given collision axis (which is transverse to the $x$-axis) and assumes that
each particle of the projectile can scatter elastically with any particle of the target.  The total rapidity $Y$ of the process is divided between
the right mover system (R), the projectile, which has rapidity $\delta Y\equiv Y-Y_0$ and the left mover system (L), the target, which has rapidity $-Y_0$. Let $P_{\rmR}[n(x_{\rmR}),Y-Y_{\rm 0}]$ and $P_{\rmL}[m(x_{\rmL}),Y_{\rm 0}]$ be the probability densities to find given configurations in the two systems, these being described as functions of
the densities of particles at the point $x$. The average $S$-matrix is given by\footnote{Here we follow the same notation used in \cite{Dumitru:2007ew}}
 \beq\label{eq:Sfact}
    \avg{S}_Y =
    \int \mcal{D}n \mcal{D}m\, P_{\rmR}[n(x_{\rmR}),Y-Y_{\rm 0}]\,
    P_{\rmL}[m(x_{\rmL}),Y_{\rm 0}]\, S[n(x_{\rmR}),m(x_\rmL)].
   \eeq
Here, $S[n(x_{\rmR}),m(x_\rmL)]$ is the $S$--matrix associated with a given
pair of configurations and the $\avg{\cdots}$ symbol represents the average
over all possible configurations $\{n(x_{\rmR})\},\{m(x_\rmL)\}$. This
'event-by-event' $S$--matrix is given by
    \beq\label{eq:Sgiven}
     S[n,m] = \exp\left[\int \dif x_\rmR \dif x_\rmL n(x_\rmR)
    m(x_\rmL) \ln \sigma(x_\rmR|x_\rmL)\right], \eeq
where $\sigma(x_{\rmR}|x_{\rmL}) = 1 - \tau(x_{\rmR}|x_{\rmL})$ is the
$S$--matrix for the scattering of two elementary particles of logarithmic
sizes $x_{\rmR}$ and $x_{\rmL}$, and $\tau(x_{\rmR}|x_{\rmL})$ the
corresponding $T$--matrix ($0 \leq \tau(x_{\rmR}|x_{\rmL})\leq 1$).

The probability densities obey the following evolution equation (the details
of the evolution can be found in references \cite{onedim,Dumitru:2007ew})
	\beq\label{eq:master} \frac{\dif P[n(x),Y]}{\dY} = \int\rmd z\,
 		f_z[n(x)-\delta(x-z)]\,P[n(x)-\delta(x-z),Y] -\int\rmd z\,
 		f_z[n(x)]\,P[n(x),Y],
 	\eeq
where $f_z[n(x)]$ is the probability per unit rapidity to find an extra
particle with logarithmic size $z$ after an evolution step (after
a small increment in rapidity, only one extra particle can be
emitted), given that the initial configuration of the evolved
system was $n(x)$. The functional form of the "deposit"
rate density $f_z[n(x)]$ can be found by assuming Lorentz invariance, and
one gets
   \beq\label{eq:fsimple}
    f_z[n(x)] = \frac{T_z[n(x)]}{\alpha(z)},
   \eeq
where $T_z[n(x)]$ is the $T$-matrix for the scattering of a particle of logarithm size $z$ off a system with a given configuration $n(x)$, and is given by
   \beq\label{eq-Tgen}
     T_z[n(x)] = 1 - \exp\left[\int \dif x\, n(x) \ln \sigma(z|x)\right],
   \eeq
and $\alpha(z)$ is the coupling parameter. In the case of running
coupling, $\alpha(z) = 1/\beta z$---in such a way to mimic the
one-loop running coupling of QCD---, with $\beta$ being the analog of
the one-loop beta function of QCD. Another important feature of
the model is the specification of the explicit form of the elementary particle-particle scattering amplitude $\tau(x|y)$, which, in analogy
with the corresponding quantity in QCD (the amplitude for dipole-dipole
scattering), is chosen as
   \beq\label{eq-tau}
     \tau(x|y) = \alpha(x) \alpha(y) \exp(-|x-y|) \equiv \alpha(x) \alpha(y)
K(x,y)\equiv 
     \alpha_x \alpha_y K_{xy}.
   \eeq
With the above expressions at hand, one can now present the evolution equations
for any observable. In particular, since we want to describe the cross section
(\ref{eq:sigmatot_3}), we will present the resulting equations for the
scattering amplitudes.

\subsection{Evolution of the amplitudes}

Let us consider a generic observable ${\cal O}$ which depends on the
configuration of the particles in the system. If one evaluates its
average value at rapidity $Y$, one gets a measurable quantity, given by
	\be \label{eq:obs}
		\avg{{\cal O}}_Y=\int{\cal D}nP[n(x),Y]{\cal O}[n(x)],
	\ee
where here we mean that the average is taken over all the configurations of the right mover, that is, $P[n(x),Y]\equiv P_R[n(x),Y]$ (the left mover consists
in a given configuration of particles).

By using Eqs.(\ref{eq:master}) and (\ref{eq:obs}), it is straightforward
to obtain the evolution equation for any physical observable:
	\be
		\frac{\partial \avg{{\cal O}}_Y}{\partial Y}=
		\int\rmd z\avg{f_z[n(x)]\left\{{\cal O}[n(x)-\delta(x-z)]
		-{\cal O}[n(x)]\right\}}_Y.
	\ee

If the observable is the amplitude for the scattering between a
projectile which consists of a single particle of a given logarithmic size $x$ and a generic target, one has (the average over $Y$ is implicit)
    \beq\label{eq:Tone}
    \frac{\del \avg{T_x}}{\del Y} =
    \alpha_x \int\rmd z\, K_{xz} \avg{T_z(1-T_x)},
   \eeq
which is not a closed equation for $\avg{T_x}$, but the first equation of
an infinite hierarchy. This equation is analogous to the first
equation of the Balitsky--JIMWLK hierarchy, which is identical to the first
of the pomeron loop (PL) equations (extended to running coupling):
it has a linear term, proportional
to $\avg{T}$, and a non--linear (quadratic) term, proportional to ${T^2}$.
The term
corresponding to the particle number fluctuations appears
only in the second equation of the hierarchy, which reads
	\beq\label{eq:Ttwo}
   \frac{\del \avg{T_x T_y}}{\del Y} \,&=&\,
   \alpha_x\int\rmd z\, K_{xz} \avg{T_z T_y (1-T_x)}+
   \alpha_y \int\rmd z\, K_{yz} \avg{T_z T_x (1-T_y)} \nn
   \,&+&\,\alpha_x \alpha_y \int\rmd z \,\alpha_z K_{xz} K_{yz}
   \avg{T_z (1-T_x)(1-T_y)},
   \eeq
the fluctuation term being the one proportional to $\avg{T}$ in the second line
of the above equation. 

In the mean field approximation (MFA), the whole hierarchy reduces to a single
closed equation, which is obtained by making $\avg{TT}=\avg{T}\avg{T}$ in
Eq.(\ref{eq:Tone}), resulting in the analogous to the (running coupling)
Balitsky--Kovchegov (BK) equation
    \beq\label{eq:T-mfa}
    \frac{\del \avg{T_x}}{\del Y} =
    \alpha_x \int\limits_z K_{xz} \left[\avg{T_z}-\avg{T_z}\avg{T_x}\right].
    \eeq
The evolution equations with fixed coupling can be straightforwardly obtained
by simply making $\alpha=$ constant. Now, we will make a brief review of the
main aspects of the evolution of the average amplitude $\avg{T}$, in both
fixed and running coupling cases.\\

\texttt{(i)} \textit{Fixed coupling case \cite{onedim}:}

\begin{enumerate}
\item From the similarity with the BK equation, Eq.(\ref{eq:T-mfa}) admits the so
called travelling wave solutions, which means that, at very large values of rapidity, the amplitude depends on $x$ and $Y$ through the scaling variable $x-x_s(Y)$, that is, the amplitude $T$ is a front which interpolates between
1 and 0 and, as $Y$ increases this front gets simply translated towards larger values of $x$, without being distorted. The function $x_s(Y)$ is the
\textit{saturation scale}, which naturally emerges from the non--linear
evolution; it separates between the dense target region, $x\simle x_s$, where $T=1$, and the dilute target region, $x\simge x_s$, where $T$ decreases exponentially. It is also an increasing function of rapidity, the analogous to the (logarithm of the) saturation momentum which emerges from the non--linear evolution in QCD, $\ln(Q_s^2/Q_0^2)$. It can be also defined as the position
of the front, that is, the line along which the amplitude is constant and of ${\cal O}(1)$ (it is usual to choose $T(x = x_s(Y ), Y ) = 1/2$). The dependence
on the combined variable $x-x_s$ is the so-called \textit{geometric scaling}
\cite{geometric,MS06}, which is valid in a window which grows with increasing
rapidity like $\propto Y^{1/2}$.

\item With the inclusion of the fluctuation effects, the differences with respect to
the mean field analysis are very significant. From a given initial condition (at $Y=0$), the evolution up to $Y$ generates a statistical ensemble of fronts,
which have the same form, but differ from each other by their respective front
positions $x_s$, and this position is now a random variable. In the fixed
coupling case, to a very good approximation, the distribution of $x_s$ with
$Y$ is a Gaussian, with both the expectation value $\avg{x_s}$ and the
dispersion $\sigma^2$ rising linearly with $Y$. The individual fronts exhibit geometric scaling, but only over a compact region, in contrast with the mean field amplitude, for which the scaling window is ever increasing with $Y$. The
average amplitude $\avg{T}$ is obtained by averaging over the ensemble, and one gets that the geometric scaling property of the individual fronts is washed out by the dispersion of the fronts, and is replaced, at sufficiently large $Y$, by
the \textit{diffusive scaling}.
\end{enumerate}

\texttt{(ii)} \textit{Running coupling case \cite{Dumitru:2007ew}:}

\begin{enumerate}
\item In the MFA, geometric scaling is also present at
asymptotic rapidities in the evolution of the amplitude, but the front formation is delayed in comparison with the fixed coupling case: the window for 
geometric scaling grows with increasing rapidity like $\propto Y^{1/6}$.

\item After the inclusion of the fluctuations, the growth of
the dispersion with $Y$ is suppressed, and one has $\sigma^2\propto\sqrt{Y}$. Besides, the influence of the fluctuations is strongly suppressed, remaining
negligible for all the rapidities of interest. In particular, the average
amplitude exhibits \textit{approximate} geometric scaling:
the deviation from geometric scaling with increasing $Y$ is too small.

\end{enumerate}

\section{Results}\label{sec:results}

Now we can study the consequences of the properties of the scattering amplitudes
discussed above on the behavior of the cross section (\ref{eq:sigmatot_3}) and
the diffractive probability (\ref{eq:Pdiff}) with increasing $Y$. Concerning
the inclusive cross section (\ref{eq:sigmatot_3}), our aim here is only to reproduce the results obtained in \cite{HIMST06} and \cite{Dumitru:2007ew},
respectively, at fixed and running coupling. The diffractive case is our main
result: in the high energy limit, it has been shown that, in the fixed coupling
case, diffractive cross section exhibits diffusive scaling \cite{HIMST06}.
Here, for the first time, we study the behavior of this quantity with increasing
energy in the presence of both fluctuation and running coupling effects.

For the purposes cited above, we must use as the input for the average scattering amplitude for particle--hadron (dipole--hadron) the solution of Eqs.(\ref{eq:Tone}) (when  fluctuations are included) and (\ref{eq:T-mfa}) (when fluctuations are not included, that is, in the MFA).
The parameters which enter into the expression for the cross section
(\ref{eq:sigmatot_3}) must be fixed: we set 
$Q_0^2=1$ GeV${}^2$, the electromagnetic coupling constant $\alpha_e=1/137$, the number of colors $N_c=3$, and only light quarks (u, d, s) enter into this analysis, with zero masses. In both inclusive and diffractive cases,
we perform the analysis first considering the fixed coupling case, in both mean field appoximation (MFA) and with fluctuations included and then we generalize it by doing the same in the case with running coupling. In the specific case of
diffractive onium--hadron scattering, in the evaluation of the probabilities
$P(\{N\}; \delta Y)$ for a given configuration of the projectile onium at rapidity $\delta Y$, as well as the averages over all the target configurations,
we follow the same procedure described in \cite{onedim}.

\subsection{Fixed coupling case}

\begin{figure}[!ht]
\centering
\scalebox{1.1}{\includegraphics*[]{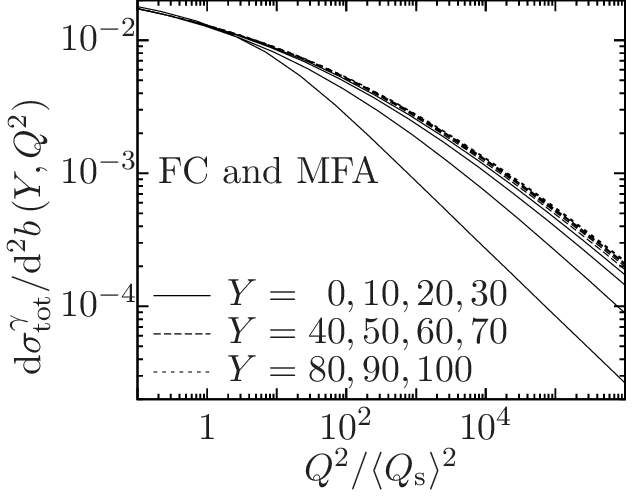}}
\scalebox{1.1}{\includegraphics*[]{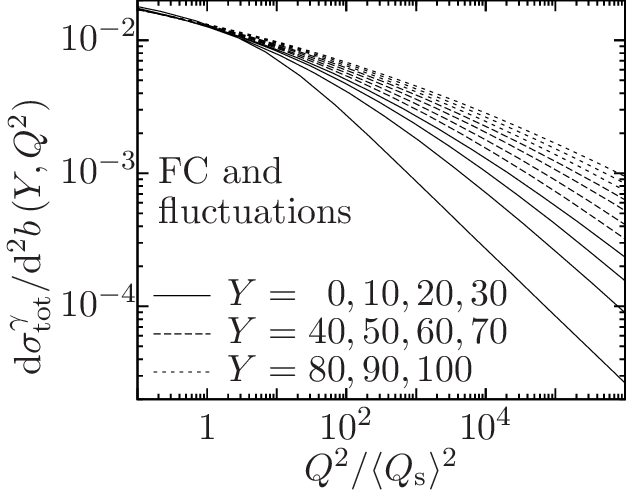}}
\caption{Fixed coupling results for various rapidities as a function of $Q^2/\langle Q_\text{s}^2\rangle$, in the MFA (left plot) and with the
inclusion of the fluctuations (right plot), up to rapidity $Y=100$.}
\label{fig:FC}
\end{figure}

\begin{figure}[!ht]
\centering
\scalebox{1.1}{\includegraphics*[]{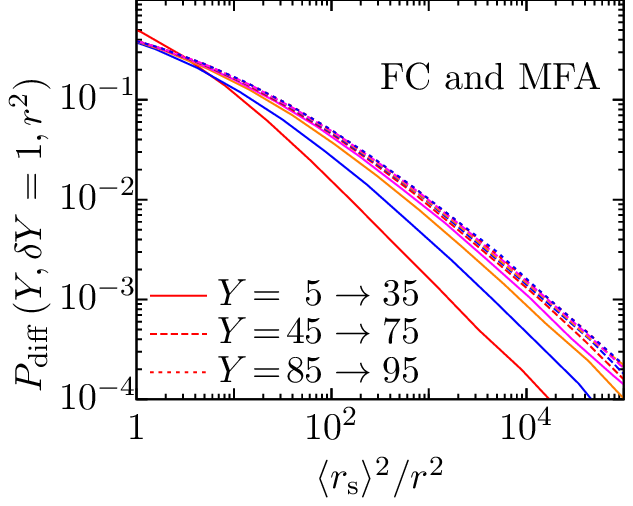}}
\scalebox{1.1}{\includegraphics*[]{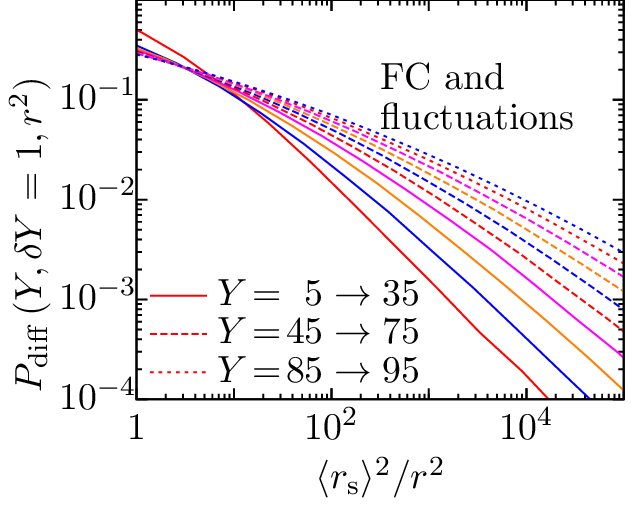}}
\caption{Diffractive probability for onium-hadron scattering as a function
of the scaling variable $\avg{r_s^2}/r^2$: fixed coupling results for various rapidities, in the MFA (left plot) and with fluctuations (right plot), up to rapidity $Y=95$.}
\label{fig:diff_FC}
\end{figure}

The results in the case of fixed coupling (FC) are shown in Figures
\ref{fig:FC} and \ref{fig:diff_FC}. The value
of the coupling constant is chosen to be $\alpha=0.2$. Fig. \ref{fig:FC} presents the DIS inclusive cross section as a function of the variable
$Q^2/\avg{Q_s^2}$, for different values of rapidity, up to $Y=100$. 
One should remember that the average saturation momentum, $\avg{Q_s^2(Y)}$, is related to the average saturation scale $\avg{x_s(Y)}$:
$\avg{x_s(Y)}=\ln(\avg{Q_s^2(Y)}/Q_0^2)=\ln(1/\avg{r_s}^2Q_0^2)$.
In the MFA (left plot), one clearly sees the geometric scaling, as well
as the growth of its window as rapidity increases from $Y=0$. For values of
rapidity values $Y\simge 30$, the curves for the cross section have the same
shape and they depend only on the scaling variable $Q^2/\avg{Q_s^2}$. After
the inclusion of the fluctuations (right plot), the curves 
deviate from the mean field behaviour (and thus from geometric scaling)
as $Y$ increases. These FC results for the inclusive cross section reflect the corresponding behaviour of the scattering amplitude and are consistent with the ones already obtained in the QCD framework \cite{HIMST06}.

In Fig, \ref{fig:diff_FC} the diffractive probability for onium-hadron DIS is
shown as a function of the geometric scaling variable $\avg{r_s}^2/r^2=e^{-(\avg{x_s}-x)}$ for different values of the total rapidity
interval $Y$. The rapidity interval of the projectile onium, $\delta Y=Y-Y_0$,
is kept fixed at a small value ($\delta Y=1$), to ensure that the projectile is a dilute system, consisting of a small number of particles (dipoles). In the MFA,
geometric scaling is reached at very large values of $Y$. When fluctuations are
included, one observes that, similarly to the inclusive cross section,
geometric scaling breaks down and $P_{diff}$ exhibits diffusive scaling. This
result is consistent with what has been found in \cite{HIMST06}

\subsection{Running coupling case}

\begin{figure}[!ht]
\centering
\scalebox{1.1}{\includegraphics*[]{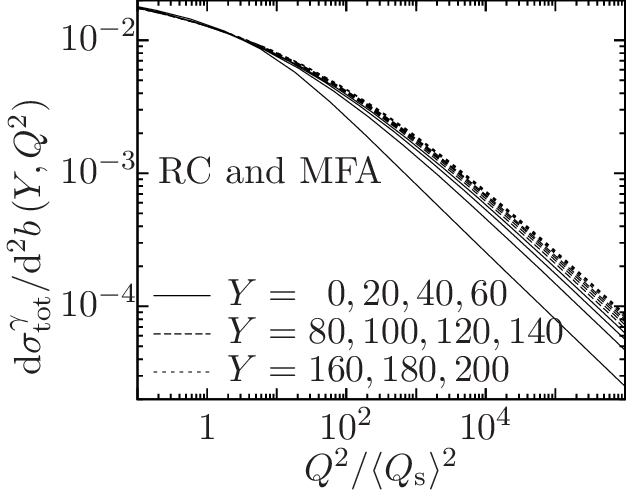}}
\scalebox{1.1}{\includegraphics*[]{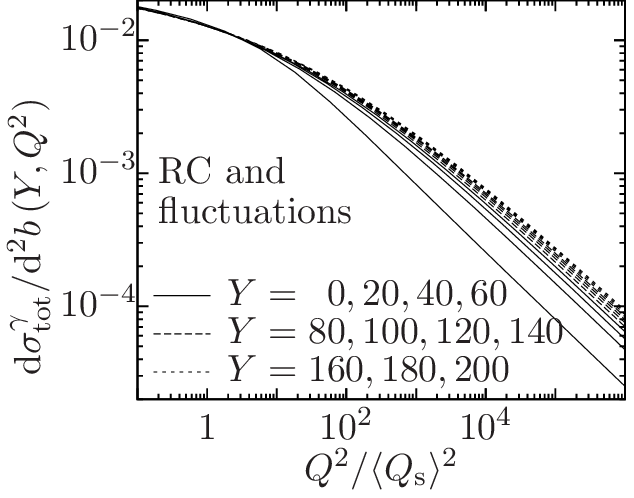}}
\caption{Running coupling results for various rapidities as a function of $Q^2/\langle Q_\text{s}^2\rangle$, in the MFA (left plot) and with the
inclusion of the fluctuations (right plot), up to rapidity $Y=200$.}
\label{fig:RC}
\end{figure}

Now, we proceed with a generalization of the previous case, by taking
into account the running of the coupling, given by $\alpha_x=1/\beta x$,
with $\beta$ chosen to be 0.72.
The results are shown in Figure \ref{fig:RC}, where the cross section (\ref{eq:sigmatot_3}) is represented 
as a function of the variable $Q^2/\avg{Q_s^2}$, for different values of rapidity, up to $Y=200$. In the MFA, one can observe geometric scaling, but
it is reached at larger values of rapidity in comparison with the FC case.
This reflects the corresponding behaviour of the scattering amplitude, for which
the formation of the front in the RC case is delayed. With the inclusion of the
fluctuations, one can observe that the increasing dispersion present in the
FC case is strongly suppressed and one has an approximate geometric scaling,
since the different curves have quite small deviations from each other when
increasing rapidity, resulting in a behavior very similar to the MFA (with
running coupling). Then, the high energy behavior of the inclusive cross section reflects the corresponding behaviour of the average particle (dipole) scattering amplitude in the running coupling case, as expected.

\begin{figure}[!ht]
\centering
\scalebox{1.1}{\includegraphics*[]{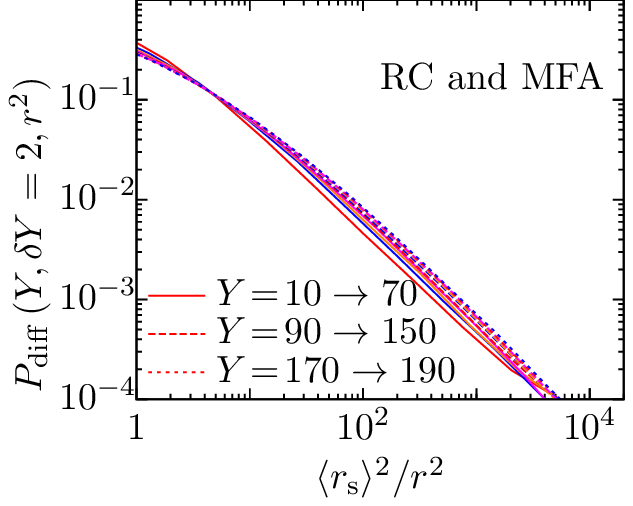}}
\scalebox{1.1}{\includegraphics*[]{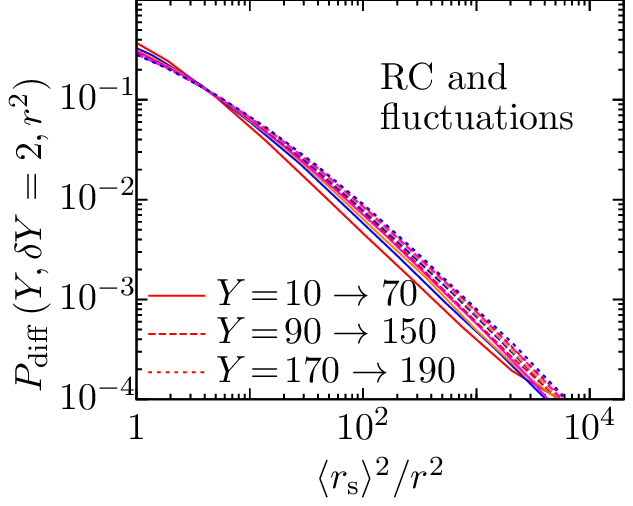}}
\caption{Diffractive probability for onium-hadron scattering as a function
of the scaling variable $\avg{r_s^2}/r^2$: running coupling results for various rapidities, in the MFA (left plot) and with fluctuations (right plot), up to rapidity $Y=190$.}
\label{fig:diff_RC}
\end{figure}

Our next step is to investigate if the suppression obtained in the inclusive case, due to both fluctuation and running coupling effects, holds also for the diffractive probability $P_{diff}(r,Y,\delta Y)$. This answer is the main result of this paper. First, from the left plot in Fig. \ref{fig:diff_RC} we can see that, in the MFA, geometric scaling is observed, as expected, but is reached faster than in the FC case, at smaller values of rapidity (now $\delta Y=2$).
Finally, in the right plot we present, for the first time, the study of the behavior of the diffractive probability in the presence of both fluctuation and RC effects. The suppression of fluctuations exists and is as strong as in the MFA case. Therefore, in diffractive DIS, within the framework of the toy model for high energy QCD, fluctuations are strongly suppressed by the running of the coupling and diffusive scaling of the cross sections, predicted in the FC case, is washed out.

\section{conclusions}\label{sec:conclusions}

In this paper we have investigated the high energy behavior of the total cross section for virtual--photon--hadron DIS and for onium--hadron diffractive DIS
within the framework of the (1+1)--dimensional model \cite{Dumitru:2007ew}, which provides a way to study, at fixed impact parameter, the effects of the particle number fluctuations and running coupling, taken into account simultaneously. In the fixed coupling
case, the results are consistent with those obtained in the framework of
QCD \cite{HIMST06}, that is, the geometric scaling which is present in
the mean field approximation at large values of rapidity, is completely washed out when fluctuations are taken into account.

By generalizing the analysis done in \cite{HIMST06}, through the inclusion of
running coupling effects, we have reproduced the results obtained in
\cite{Dumitru:2007ew} for the inclusive virtual photon--hadron cross section:  the behaviors of this cross section with and without fluctuations are similar, this observable presenting approximate geometric scaling, which means that
the running of the coupling suppresses the fluctuation effects at asymptotic
rapidities. In the diffractive onium--hadron scattering, the diffractive
probability exhibits geometric scaling in the MFA. When fluctuations
are included, diffusive scaling is seen in the fixed coupling case,
while that, in the running coupling case, geometric scaling is present
and reached at smaller values of rapidity $Y$ than in the case without fluctuations.

This suggests that the mean field treatment with running coupling
would be enough to study not only the inclusive lepton-hadron DIS, but
also the diffractive DIS, for all the energies available at present and to be available in a near future. The toy model also allows the investigation of the other processes which, in the framework of QCD, admit a dipole factorization. Thus, it would be interesting to apply it to such processes, in particular
less inclusive ones, in order to investigate if the suppression of fluctuations by running coupling effects remains present.

\section*{Acknowledgments}
We would like to thank Andrea Beraudo and Yuri Kovchegov
for very useful discussions. MBGD akcnowledges the hospitality of the
CERN Theoretical Division and JTSA acknowledges the hospitality of
GFPAE-IF-UFRGS. This work has been supported by CNPq and FAPERGS
(Brazil). E.G.O.\ is supported by FAPESP (Brazil) under contract
2011/50597-8.

\bibliographystyle{unsrt}
\bibliography{myrefs}
\end{document}